\documentclass[a4paper,11pt,twoside,onecolumn]{article}
\usepackage{amsmath}
\usepackage[affil-it]{authblk}
\usepackage[bottom=3cm,top=3cm,left=3cm,right=3cm]{geometry}
\usepackage{epsf}
\usepackage{latexsym}
\usepackage{amsbsy}
\usepackage{amssymb}
\usepackage{amsfonts}
\usepackage{bbm}
\usepackage[T1]{fontenc}
\usepackage{textcomp}
\usepackage{helvet}
\usepackage{ifthen}
\usepackage{color}
\usepackage{graphicx}
\usepackage{eso-pic}
\usepackage{multirow}
\usepackage{appendix}

\usepackage{cite}

	\newcommand{\be}{\begin{equation}}
	\newcommand{\ee}{\end{equation}}
	\newcommand{\ben}{\begin{equation*}}
	\newcommand{\een}{\end{equation*}}

	\newcommand{\vb}[1]{\ifmmode\mathbf{#1}\else$\mathbf{#1}$\fi} 

	\newcommand{\argg}[2]{(#1 , #2)}
	
	\newcommand{\azul}[1]{\textcolor{black}{#1}}
	
	\newcommand*{\Scale}[2][4]{\scalebox{#1}{$#2$}}%

\begin{document}
\title{Acoustic scattering by two fluid confocal prolate spheroids}
\author{E. F. Lavia}
\maketitle
\begin{abstract}
The exact spheroidal-function series solution for the time-harmonic acoustic scattering of a plane wave by
two fluid confocal prolate spheroids is developed and a numerical implementation is formulated and validated
by independent methods.
The two spheroids define three regions in which the acoustic fields are expanded in terms of spheroidal wave functions
multiplied by unknown coefficients. 
These expansions are forced to satisfy the boundary conditions and by using the orthogonality properties of the 
involved functions an infinite matricial system for the coefficients is obtained. The resulting system is then 
solved through a truncation procedure. 
The implementation has no limitations regarding the sound speed and density of the three media involved or in
the incidence frequency. 
\end{abstract}

\section{Introduction}
\label{intro}

The availability of exact solutions for certain acoustic scattering problems (involving simple geometries as spheres, cylinders,
etc.) has been, besides its importance {\it per se}, of widespread utility since additionally they can be used as benchmark solutions 
to validate approximate but more general methods based on some kind of discretization (being the Finite Element Method --FEM-- and the 
Boundary Element Method --BEM-- maybe the most prominent examples).
Furthermore, exact solutions usually are also less prone to display limitations or problems in the high frequency regime.

In the case of acoustic scattering of harmonic plane waves by obstacles with those simple geometries, exact solutions 
found by using the separation-of-variables procedure exist when the obstacle surface can be identified with a coordinate surface
that belongs to a coordinate system for which the Helmholtz equation is separable \cite{morse1948vibration}.
Restricting to second degree surfaces, the latter condition corresponds to eleven coordinate systems of which the spherical, 
cylindrical and prolate/oblate spheroidal surely are the most conspicuous \cite{moon1971field} because they are the ones that best
fit in relevant scattering situations.

Solutions for the scattering produced by the infinite circular fluid cylinder or the fluid sphere \cite{anderson1950sound} are
{\it classics} and usually conform the starting point of all introductory texts about penetrable acoustic scattering.
The corresponding solutions for the acoustic elastic problem, when there are also shear waves in addition to compressional ones,
appeared shortly afterwards \cite{faran1951sound,hickling1962analysis}.

The same scheme used to obtain the solution for the scattering by a single obstacle (that is matching boundary conditions on a coordinate
surface) can be used to build the solution for the scattering of two or more similar obstacles, each one inside the previous, 
because in that case the boundary conditions have to be verified in coordinate surfaces of the same type. 
In the spherical coordinate system, for example, that procedure leads naturally to the solution for the scattering by two concentric 
spheres (a setup which is called a {\it spherical shell}).
It is important to remark that the method provides a straightforward solution only if the spheres have the same origin, i.e. if
they are concentric, because otherwise the boundary conditions do not correspond to evaluate the solution in a single coordinate surface.

Seminal studies on spherical and cylindrical shells appeared in the sixties \cite{goodman1962reflection,hickling1964analysis,doolittle1966sound}.
Today these results are firmly set within {\it the canon} of highly verified acoustic scattering solutions.
Subsequently many other works have considered them under different conditions \cite{mcnew2009sound,everstine1998acoustic,cummer2008scattering}. 



The prolate and oblate spheroids, whose geometries make them susceptible to many practical acoustical
applications, have been object of much interest
\cite{spence1951scattering,varadan1982computation,roumeliotis2007acoustic}.
Some of the special functions resulting from the separation of variables of the wave equation in these 
coordinate systems (the spheroidal wave functions \cite{morse-feshbach,flammer1957,skud1971}) display in
their calculation difficulties greater than those corresponding to the spherical and cylindrical cases 
(spherical and cylindrical Bessel functions, respectively) which is why they usually require numerical
precision beyond the current 64-bit hardware precision \cite{van2004improved,falloon2003theory}. 
The history of its calculation is very prolific, see \cite{van2004improved,falloon2003theory,agd2014}
and references therein.
\azul{
An approach that circumvents the difficulties associated with spheroidal functions is the one based
on the Vekua transformation \cite{charalam2002}, which connects the kernels of Laplace and Helmholtz
equations, thus providing an analytical solution for the scattering by spheroids without using spheroidal
functions. However, the numerical implementation of this approach still requires arbitrary precision 
arithmetic \cite{gergidis2007,gergidis2018}.
}

The exact analytical solution for the fluid spheroid appeared in 1964 \cite{yeh1964}.
Afterwards, different works dealing with numerical calculations were restricted to: certain particular 
cases of sound speed or density contrasts \cite{yeh1967,prario2015prolate}, the low frequency regime 
\cite{burke1968scattering,einspruch1961scattering,yeh1967,furusawa1988prolate,ye1997low,tang2009average} 
or low eccentricity spheroids \cite{kotsis2008acoustic}.

A numerical evaluation of the exact solution without any limitation, based on \cite{yeh1964} and using 
a computational code \cite{agd2014} for spheroidal wave function calculation in arbitrary precision was
presented in \cite{gonzalez2016computational}.

A configuration with two spheroids was addressed in \cite{yeh1969scattering}, where scattering of a plane
wave by a rigid prolate spheroid coated with a confocal sheat of penetrable (fluid) acoustic material was 
obtained. 
A system of multilayered confocal prolate spheroids, the innermost being considered rigid, was developed 
in \cite{charalambopoulos2001scattering} and then applied to the case of an spheroid coated with a single
layer of fluid \cite{charalambopoulos2002scattering}, providing thus a simplified model for a stone located
in the human kidney. In all these works, the interior spheroid is always considered as an impenetrable one.

The elastic prolate spheroid was addressed in \cite{silbiger1963scattering}. 
In that reference, the scattering from a prolate spheroidal 
shell was approximated by the response due to one in a resonant mode. 
Approximations for the scattering by spheroidal elastic shells in the resonance 
region and calculated with the T-matrix method are presented in \cite{gaunaurd1990acoustic}.

This work presents a numerical evaluation of the {\it exact}, in terms of a series, solution for the acoustic 
scattering of two fluid prolate confocal spheroids (from now on this setup will be called a {\it spheroidal shell}) 
valid for any value of eccentricity and arbitrary fluid properties of the three involved physical mediums.
The oblate case can be worked out following the same lines with only slight modifications, see for details
\cite{gonzalez2016computational}. In view of that, this work is devoted to the prolate spheroid.
The numerical implementation was developed using a modified version of the computational codes by Adelman et al. \cite{agd2014}.

This paper is structured as follows. In Section \ref{theory} the analytical solution for the acoustic scattering by the 
spheroidal shell is formulated. Section \ref{num_impl} provides the workings of the numerical implementation. In Section \ref{verifications}
several numerical verifications against certain limiting cases (spheroid tending to sphere) and with results provided by a
BEM implementation are carried out. Computations of external and internal fields are also included.
Conclusions of the work are summarized in Section \ref{conclusions}.

\section{Theory. Analytical solution}
\label{theory}

The time-harmonic acoustic scattering of a plane wave by an spheroidal shell can be solved, as said previously, by separating 
variables in prolate spheroidal coordinates $(\xi,\eta,\varphi)$ \cite{morse-feshbach}. These coordinates are defined by
\begin{equation}
\left \{ 
\begin{matrix} 
\displaystyle x = \frac{d}{2} \; [(\xi^{2}-1 ) (1-\eta^{2})]^{1/2} \cos {\varphi}\\
\\
\displaystyle y = \frac{d}{2} \; [(\xi^{2}-1 ) (1-\eta^{2})]^{1/2} \sin {\varphi}\\
\\
\displaystyle z = \frac{d}{2} \; \xi\hspace{2pt} \eta,
\end{matrix}\right.
\label{spheroidal_coord}
\end{equation}
where $d$ is the interfocal distance of the ellipse of major semi-axis $a=(d/2) \hspace{2pt} \xi$ and minor 
semi-axis $b=(d/2) \hspace{2pt} (\xi^{2}-1 )^{1/2} $. 
The values for the prolate spheroidal coordinates must verify $ \xi \geq 1, -1\leq \eta \leq 1,$ and $0 \leq \varphi <2\pi$.
The parameter $ d = 2 (a^2 - b^2)^{1/2} $ defines a particular prolate spheroid system. The surface of any spheroid belonging 
to this system coincides with the coordinate surface given by $\xi = \xi_0$, with $\xi_0 = (1 - (b/a)^2 )^{-1/2}$.

\begin{figure}[!htb]
	\centering
	\includegraphics[scale=0.30]{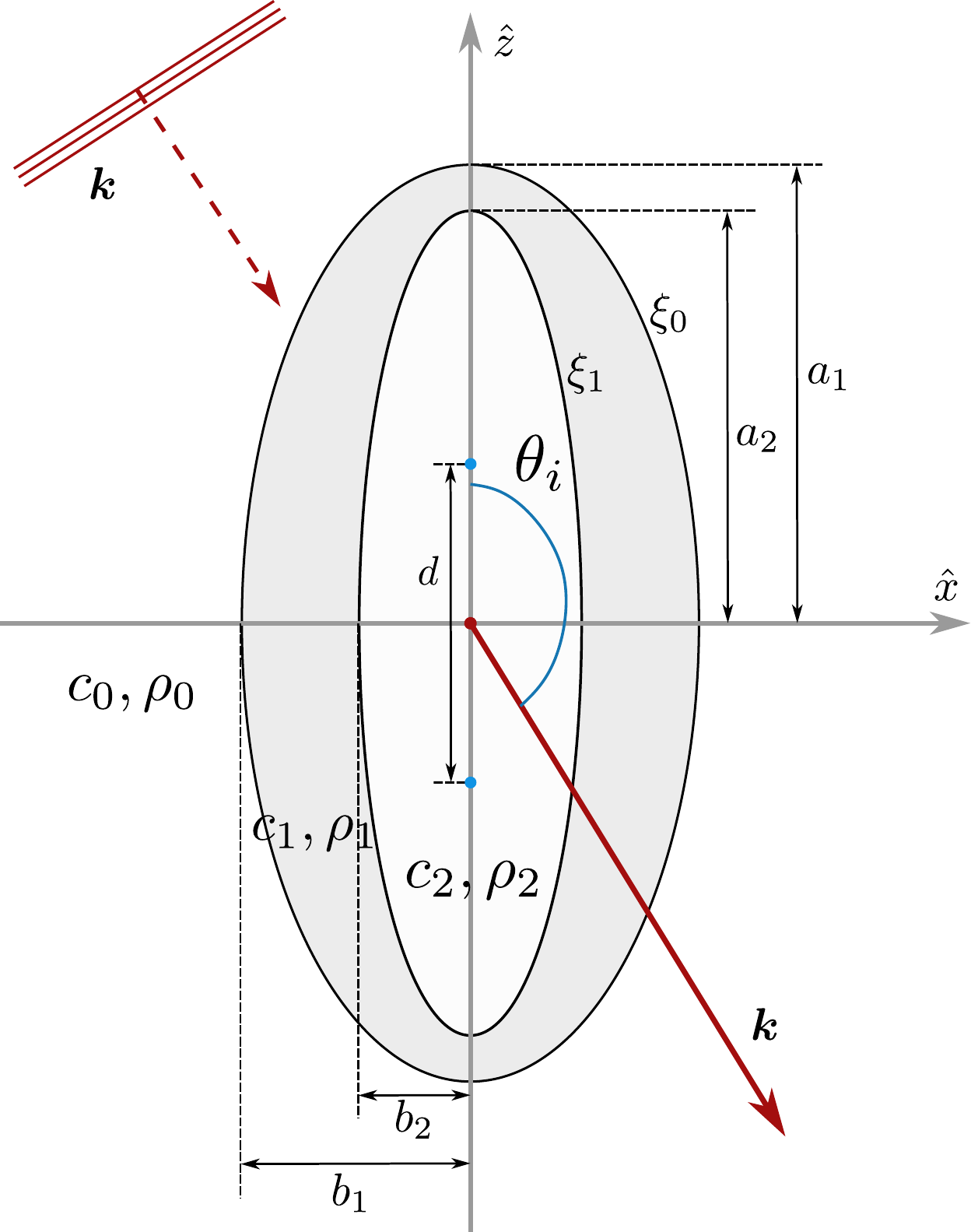}
	\caption{Coordinates for the scattering of a plane wave with incidence wave vector $\boldsymbol{k}$ by two 
	confocal prolate spheroids.}
	\label{fig_scheme}
\end{figure}

The scattering problem is depicted in Figure \ref{fig_scheme}.
The acoustic pressure of an incident plane wave with angular frequency $\omega$, propagating in a surrounding medium of 
sound speed $c_0$ can be written as
\begin{equation*}
	p_i =  p_0 \exp( i k_0 \hat{k} \cdot \boldsymbol{x} ),
\end{equation*}
where $ k_0 = \omega/c_0 $ is the wave number, $\hat{k} = (\sin\theta_i \cos\varphi_i, \sin\theta_i \sin\varphi_i, 
\cos\theta_i)$ is the incidence direction (being $\theta_i,\varphi_i$ the spherical angles of incidence) and $p_{0}$ 
the amplitude.
Without loss of generality, due to the symmetry of revolution around the $z$ axis, it can be considered $\varphi_i=0$ 
so that $\hat{k} = (\sin\theta_i , 0 , \cos\theta_i)$ and the incidence is fully characterized by a single angle. 
Such incident wave on the prolate spheroidal shell is illustrated in Figure \ref{fig_scheme} and identified with the wave vector 
$\boldsymbol{k} = k_0 \hat{k}$.

The two spheroids constituting the shell have major and minor semiaxis $a_1$, $b_1$ and $a_2$, $b_2$, respectively, 
and verify the condition 
\begin{equation}
        a_1^2 - b_1^2 = a_2^2 - b_2^2 = \left( \frac{d}{2} \right)^2,
        \label{confocal_relations}
\end{equation}
which assures that the focal distance $d$ is the same for both (i.e. the spheroids are confocal). Therefore, both 
spheroids are described by the same spheroidal system, related to cartesian coordinates by \eqref{spheroidal_coord}.
The boundaries of the spheroids correspond to values $\xi_0$ and $\xi_1$ of the spheroidal coordinate $\xi$ 
and define three regions characterized by different values of sound speed and density $c_i,\rho_i$ ($i=0,1,2$).

The procedure of separation of variables applied on the Helmholtz equation $ (\nabla^2 + k^2) p = 0 $ in the coordinates $(\xi,\eta,\varphi)$ 
leads to a representation of the solution in terms of spheroidal angular functions $S_{mn}(h,\eta)$ and radial spheroidal 
functions of the first and second kinds, $R_{mn}^{(1)}(h,\xi)$ and $R_{mn}^{(2)}(h,\xi)$, respectively 
\cite{morse-feshbach,flammer1957,skud1971}.
These functions also depend on the dimensionless parameter $ h \equiv (d/2) \: k $, which characterizes the scattering in 
each medium through its corresponding wave number $k$.

Then, in each of the three regions a Helmholtz equation with a different wave number $k_i = \omega / c_i, (i=0,1,2)$ is valid so that the fields there
are built of linear combinations of the $S_{mn}(h_i,\eta), R_{mn}^{(1)}(h_i,\xi)$ and $R_{mn}^{(2)}(h_i,\xi)$ spheroidal wave functions
with unknown coefficients.
The continuity of the pressure and normal velocity at each boundary $\xi_0, \xi_1$ leads to a system of matrix equations; 
since in this case there are four (two conditions times two boundaries) equations, then four matrix unknowns are expected.

In the following paragraphs the continuity equations will be transformed into a matrix system which will allow to solve the
scattering problem for the spheroidal shell. The notation to be used closely follows the previous work \cite{gonzalez2016computational}.

In the surrounding medium $(c_0,\rho_0)$ the pressure is the sum of the incident pressure $p_i$ and the scattering pressure $p_s$, 
\be
	p = p_i + p_s.
	\label{pressure_out}
\ee
The incident pressure $p_i$ can be expanded on prolate spheroidal functions \cite{flammer1957} and written as
\ben
	p_{i} = 2 p_0 \sum_{m, n \geq m} i^n \epsilon_m \:
		S_{mn}\argg{h_0}{ \cos \theta_i } \: S_{mn}\argg{h_0}{\eta} \: R_{mn}^{(1)}\argg{h_0}{\xi} \cos(m\varphi),
\een
where $\epsilon_m$ is the Neumann factor, defined as $\epsilon_m = 2$  if $m \neq 0$ and $\epsilon_m = 1$ if $m=0$.
The $S_{mn}$ functions are assumed to be normalized, thus
\[
	\int_{-1}^{1} \:[S_{mn}(h_i,\eta)]^2 \: d\eta = 1.
\]

The scattered pressure field $p_s$ can also be expressed as linear combination of spheroidal wave functions but using 
the radiating function $ R^{(3)} \equiv R^{(1)} + i R^{(2)} $ (which diverges at $ \xi = 1 $) instead of the regular $ R^{(1)} $;
then
\be
	p_s = 2 p_0 \sum_{m, n \geq m} i^n A_{mn} \: \epsilon_m \: S_{mn}\argg{h_0}{\cos \theta_i } 
		\: S_{mn}\argg{h_0}{\eta} \: R_{mn}^{(3)}\argg{h_0}{\xi} \: \cos(m\varphi),
	\label{pscatt}
\ee
where $ A_{mn} $ is a matrix of expansion coefficients.

The pressure field inside the outer spheroid, which constitutes the $(c_1,\rho_1)$ medium, can be written as a sum of a standing
solution and a radiating solution, i.e.
\be
	p_1 = p_{1s} + p_{1r}
\ee
where
\be
p_{1s} = 2 p_0 \sum_{m, n \geq m} i^n  B_{mn} \: \epsilon_m \: S_{mn}\argg{h_0}{\cos \theta_i }
		\: S_{mn}\argg{h_1}{\eta} \: R_{mn}^{(1)}\argg{h_1}{\xi} \: \cos(m\varphi),
	\label{p1std} 
\ee
\be
 p_{1r} = 2 p_0 \sum_{m, n \geq m} i^n C_{mn} \: \epsilon_m \: S_{mn}\argg{h_0}{\cos \theta_i } 
		\: S_{mn}\argg{h_1}{\eta} \: R_{mn}^{(3)}\argg{h_1}{\xi} \: \cos(m\varphi).
	\label{p1rad}
\ee

Finally, the pressure in the inner spheroid (which must be regular in $\xi=1$) can be written as a standing wave,
\be
p_2 = 2 p_0 \sum_{m, n\geq m} i^n  D_{mn} \: \epsilon_m \: S_{mn}\argg{h_0}{\cos \theta_i }
		\: S_{mn}\argg{h_2}{\eta} \: R_{mn}^{(1)}\argg{h_2}{\xi} \: \cos(m\varphi).
	\label{p2} 
\ee

In the expressions \eqref{p1std}, \eqref{p1rad} and \eqref{p2} the matrices $ B_{mn}, C_{mn}, D_{mn} $ \azul{contain} the 
corresponding coefficients to be determined.

In each interface, $\xi_0$ and $\xi_1$, the boundary conditions of continuity of pressure and normal velocity must be \azul{fulfilled}.
This leads to the following four equations:
\be
	( p_i + p_s )|_{\xi=\xi_0} = ( p_{1s} + p_{1r} )|_{\xi=\xi_0}
	\label{bc_presion_0}
\ee
\be
	\frac{1}{\rho_0} \left. \frac{\partial( p_i + p_s )}{\partial \xi} \right|_{\xi=\xi_0} =
	\frac{1}{\rho_1} \left. \frac{\partial( p_{1s} + p_{1r} )}{\partial \xi} \right|_{\xi=\xi_0}
	\label{bc_derivada_0}
\ee
\be
	( p_{1s} + p_{1r} )|_{\xi=\xi_1} = p_2 |_{\xi=\xi_1}
	\label{bc_presion_1}
\ee
\be
	\frac{1}{\rho_1} \left. \frac{\partial( p_{1s} + p_{1r} )}{\partial \xi} \right|_{\xi=\xi_1} =
	\frac{1}{\rho_2} \left. \frac{\partial( p_2 )}{\partial \xi} \right|_{\xi=\xi_1}
	\label{bc_derivada_1}
\ee

In order to build the matrix system it is convenient, for the first two previous equations, to expand $S_{mn}( h_0,\eta)$ in terms 
of the set $\{ S_{m\ell}(h_1,\eta) : \ell \geq m \}$ and, for the last two ones, $S_{mn}( h_1,\eta)$ in terms of the set 
$\{ S_{m\ell}(h_2,\eta) : \ell \geq m \}$.
Then,
\be
	S_{mn}(h_0,\eta) = \sum_{ \ell = m }^\infty \alpha^{(m)}_{n\ell} S_{m\ell}(h_1,\eta) \qquad  \text{ with } \qquad
	\alpha^{(m)}_{n\ell} =  \int_{-1}^1 \: S_{mn}(h_0,\eta) S_{m\ell}(h_1,\eta) \: d\eta
	\label{S_expansion_0}
\ee
and 
\be
	S_{mn}(h_1,\eta) = \sum_{ \ell = m }^\infty \tilde{\alpha}^{(m)}_{n\ell} S_{m\ell}(h_2,\eta) \qquad \text{ with } \qquad
	\tilde{\alpha}^{(m)}_{n\ell} = \int_{-1}^1 \: S_{mn}(h_1,\eta) S_{m\ell}(h_2,\eta) \: d\eta.
	\label{S_expansion_1}
\ee

Substituting the expansion \eqref{S_expansion_0} in the LHS of Eqs. \eqref{bc_presion_0} and \eqref{bc_derivada_0}, and the expansion
\eqref{S_expansion_1} in the LHS of Eqs. \eqref{bc_presion_1} and \eqref{bc_derivada_1} and using the orthogonality properties of the 
families $\left\{S_{mn}(h_1,\eta)\cos(m\varphi): m\geq 0, n\geq m \right\}$ and 
$\left\{S_{mn}(h_2,\eta)\cos(m\varphi): m\geq 0, n\geq m \right\}$, four matrix equations involving the $A_{mn}, B_{mn}, C_{mn}$ and $D_{mn}$
coefficients are obtained.

It is convenient to define matrices 
\[
	[ \: Q^j ( h, \xi ) \: ]_{\:\sigma n}^{(m)} = i^n \: \alpha_{n\sigma}^{(m)} \: S_{mn}^{h_0} \: R^{(j)}_{mn}( h, \xi)
	\qquad j=3
\]
\[
	[ \: Q^{j \:\prime}( h, \xi ) \: ]_{\:\sigma n}^{(m)} = i^n \: \frac{\rho_1}{\rho_0} \: \alpha_{n\sigma}^{(m)} \: 
	S_{mn}^{h_0} \: R^{(j)\:\prime}_{mn}( h, \xi)
	\qquad j=3
\]
\[
	[ \: D^j( h, \xi ) \: ]_{\:\sigma n}^{(m)} =  i^n \: \delta_{n\sigma} \: S_{mn}^{h_0} \: R^{(j)}_{mn}( h, \xi )
	\qquad \qquad j=1,3
\]
\[
	[ \: D^{j \: \prime}( h, \xi ) \: ]_{\:\sigma n}^{(m)} =  i^n \: \delta_{n\sigma} \: S_{mn}^{h_0} 
	\: R^{(j)\:\prime}_{mn}( h, \xi ) \qquad \qquad j=1,3
\]
\[
	[ \: \tilde{Q}^j ( h, \xi ) \: ]_{\:\sigma n}^{(m)} = i^n \: \tilde{\alpha}_{n\sigma}^{(m)} \: S_{mn}^{h_0} \: R^{(j)}_{mn}( h, 
\xi)
	\qquad j=1,3
\]
\[
	[ \: \tilde{Q}^{j \:\prime}( h, \xi ) \: ]_{\: \sigma n}^{(m)} = i^n \: \frac{\rho_2}{\rho_1} \: \tilde{\alpha}_{n\sigma}^{(m)} \: 
	S_{mn}^{h_0} \: R^{(j)\:\prime}_{mn}( h, \xi)
	\qquad j=1,3,
\]
and vectors
\[
	F_\sigma^{(m)} = \sum_{n=m}^\infty i^n \: S_{mn}^{h_0} \: \alpha_{n\sigma}^{(m)} \: R_{mn}^{(1)}(h,\xi)
\]
\[
	G_\sigma^{(m)} = \sum_{n=m}^\infty \frac{\rho_1}{\rho_0} \: i^n \: S_{mn}^{h_0} \: \alpha_{n\sigma}^{(m)} \: 
	R_{mn}^{(1)\:\prime}(h,\xi)
\]
where the prime indicates the $\xi$-derivative (i.e. $^\prime \equiv d/d\xi$), $\delta_{n\sigma}$ is the Kronecker delta and 
$ S_{mn}^{h_0} \equiv S_{mn}(h_0,\cos \theta_i) $. The $D^j$ and $D^{j \: \prime}$ matrices contains $\delta_{n\sigma}$, thus they
are diagonal. This property is stressed by the ``$D$-letter'' in his name. Care must be taken in avoiding to mix those matrices with 
the unknown coefficients $D_{mn}$.

Then, it can be shown that for each fixed $ m = 0, 1, 2,... $ the four matrix equations set leads to a infinite matrix system 
\be
	\Scale[0.875]{
	\begin{pmatrix}
        Q^3(h_0,\xi_0)^{(m)} & -D^1(h_1,\xi_0)^{(m)} & -D^3(h_1,\xi_0)^{(m)} & 0  \\
        \\
        Q^{3\prime}(h_0,\xi_0)^{(m)} & -D^{1\prime}(h_1,\xi_0)^{(m)} & -D^{3\prime}(h_1,\xi_0)^{(m)} & 0  \\
        \\
        0 & \tilde{Q}^1(h_1,\xi_1)^{(m)} & \tilde{Q}^3(h_1,\xi_1)^{(m)} & -D^1(h_2,\xi_1)^{(m)}  \\
        \\
        0 & \tilde{Q}^{1\prime}(h_1,\xi_1)^{(m)} & \tilde{Q}^{3\prime}(h_1,\xi_1)^{(m)} & -D^{1\prime}(h_2,\xi_1)^{(m)} 
        \end{pmatrix}  
        \begin{pmatrix}
        A^{(m)} \\ \\ B^{(m)} \\ \\ C^{(m)} \\ \\ D^{(m)}
        \end{pmatrix} =
        \begin{pmatrix}
        -F^{(m)} \\ \\ -G^{(m)} \\ \\ 0 \\ \\ 0  
        \end{pmatrix}
        }. 
        \label{matricial_system}
\ee

The index $m$ was indicated as a superscript to emphasize the fact that for each fixed $m$ a matrix system of the 
type \eqref{matricial_system} has to be solved.
Each of these solutions provides a vector including the four coefficients $A^{(m)}, B^{(m)}, C^{(m)}, D^{(m)}$,
which contain all the corresponding $n$-values ($n = m, m+1, ...$) for that index $m$.
Once the coefficients for all $m$ have been obtained, the fields $p$, $p_1$ and $p_2$ in each region can be evaluated.

In the far-field limit it can be shown \cite{yeh1964} that, with respect to spherical coordinates $(r,\theta,\varphi)$ 
of the observation point, the scattering pressure $p_s$ is given by
\[
	p_s( r,\theta,\varphi) \approx p_0 \frac{e^{ik_0 r}}{r} f_\infty(\theta, \varphi),
\]
where $f_\infty(\theta, \varphi)$ is the so-called far-field scattering amplitude function which is widespreadly used 
in different acoustic scattering applications. In the particular case of an spheroidal shell, it results
\be
	f_\infty(\theta, \varphi)= \frac{2}{i k_0}  \sum_{m, n\geq m} A_{mn} 
	\epsilon_m \: S_{mn}\argg{h_0}{\cos \theta_i }S_{mn}\argg{h_0}{\cos \theta } \cos(m\varphi).
	\label{finf}
\ee

\section{Numerical implementation}
\label{num_impl}

To numerically calculate the set 
of coefficients $\{ A, B, C, D \}$ a truncation procedure must be carried out.
The first step is to select a maximum value $ M $ for the index $ m $.
Then, $M+1$ matricial systems \eqref{matricial_system} labeled with a distinct value $m$ results.
The system $m=0$ has $M+1$ unknowns $ A^{(0)}_n $ ($ n = 0, 1,...,M $) and, equivalently, the same number of the other coefficients.
By considering that the subsequent matricial systems $ m=1, m=2, ..., m=M$ have the same size, this leads to $4(M+1)^2$ unknowns
$ A^{(m)}_n, B^{(m)}_n, C^{(m)}_n, D^{(m)}_n $ where the label $n$ takes the values $n=m,m+1,...,m+M$.

In summary, each matricial system will have a size  $ 4 ( M + 1 ) \times 4( M + 1 )$ and its solution will provide the corresponding
$m$-set of $4(M+1)$ coefficients $A^{(m)}_n$, $B^{(m)}_n$, $C^{(m)}_n$ and $D^{(m)}_n$ with $n \in [m,m+M]$.

For example, in an hypothetical case of $ M = 5 $ there will be six systems \eqref{matricial_system} of size $36 \times 36$, each
one identified by $ m = 0,1,...,5 $. The solution of any $m$-system provides a vector
\[
	( A_0^m, A_1^m, ..., A_5^m, B_0^m, B_1^m, ..., B_5^m, C_0^m, C_1^m, ..., C_5^m, D_0^m, D_1^m, ..., D_5^m).
\]
Finally, the $6 \times 6$ matrix of coefficients $A_n^m$ results in 
\[
	A = \begin{pmatrix}
	   A^0_{\:0} & A^1_{\:1} & ... & A^5_{\:5} \\
	   A^0_{\:1} & A^1_{\:2} & ... & A^5_{\:6} \\
	   ... & \\
	   A^0_{\:5} & A^1_{\:6} & ... & A^5_{\:10} \\
	   \end{pmatrix},
\]
where the $m$-th column comes from the numerical solution of the $m$-system. The other coefficient matrices $B,C,D$ can be
arranged in a similar fashion.

The key idea in this truncation procedure is that for some truncation number $M$ the obtained values for all no negligible
coefficients should not change; then, a subsequent increase in the size of the system (i.e. a new truncation number greater than $M$)
must not alter the solution.

Convergence for a certain coefficient in the system is achieved when an increment in the truncation number $ M $ does not
appreciably change the coefficient value. 
As stated above, convergence for a particular acoustic problem will be achieved when the successive coefficients that appear as a 
consequence of considering \azul{larger} $ M $ values not lead to a appreciable change in the numerical solution.
Indicators of the occurrence of that situation usually will be coefficients tending to zero.
However, to avoid false identifications, care must be taken in selecting the appropriate $ M $ and avoid falling in stagnation zones
where the coefficients are small for certain $m,n$ but rise for $m,n$ greater. 

In the $M=5$ example provided above, the subsequent approximation $M=6$ involves thirteen new coefficients $A^0_6, A^1_7,..., 
A^6_{12}, A^6_6, A^6_7, ... , A^6_{11}$.
Even if all of them were negligible, it could happen that within the subsequent coefficients appearing for $M = 7$, for example, 
some were not negligible and, therefore, necessary to obtain a convergent solution. In any case, a calculation for excess in the
coefficients is mandatory as well as a good habit.

Coefficients of negligible value are not, however, the sole indicator of convergence because it could happen that these coefficients
were multiplied in the expansion by spheroidal functions that may take very high values for particular choices of their arguments.
Again, to watch over the emergence of this type of pathological behavior is another good habit.

\section{Verifications}
\label{verifications}

\subsection{Spherical shell}

In the limiting cases of $b_i$ tending to $a_i$ ($i=1,2$) the two confocal spheroids tend to conform a spherical shell.
Since the scattering by a spherical shell has exact solution, expressed in terms of spherical Bessel functions, it can be compared
with the scattering resulting from a spheroidal shell at this geometrical limit.

In the spheroidal system $a_i=b_i$ is a prohibited value because in that case $ d = 0 $ and the system \eqref{spheroidal_coord} 
becomes singular but nothing precludes to use $b_i$ values near its corresponding $a_i$ values and consequently considering an 
approximated spherical shell. 
For the validity of the confocal spheroidal shell model the values $a,b$ must verify the relation \eqref{confocal_relations}.
Fixing arbitrarily $a_1=0.5$, $a_2=0.25$ and $b_1=0.4999$, according to \eqref{confocal_relations} the remaining $b_2$ has the value 
$b_2=0.24979993995195438$. Using these values the farfield angular pattern $|f_\infty|$ for the spheroidal shell is compared against the 
exact spherical solution for the frequencies $f = 5, 10, 15 $ kHz and $c_i,\rho_i$ parameters according to Table \ref{data_table}
(typical values in underwater acoustics applications). 
\azul{Since the incidence angle can be chose freely because the assumed isotropy, $\theta_i = 180^\circ$ 
(equivalent to $\eta=-1$) is employed for later convenience.}

\begin{table}[h]
	\begin{center}
	\begin{tabular}{ccc}
	\hline
	\hline 
	\vspace{-0.8em}\\
	Medium & $c$ (m s$^{-1}$) & $\rho$ (kg m$^{-3}$) \\
	\hline
	\vspace{-1em}\\
	0 (water) & 1477.4 & 1026.8 \\
	1 & 1.04 $c_0$ & 1.04 $\rho_0$ \\
	2 & 0.23 $c_0$ & 0.00129 $\rho_0$ \\
	\vspace{-1em}\\
	\hline
	\hline
	\end{tabular}
	\end{center} 
	\caption{Material properties (sound speed $c$ and density $\rho$) 
	for the spherical and spheroidal shell acoustic scattering problem.}
	\label{data_table}
\end{table}

\begin{figure}[!ht]
	\centering
	\includegraphics[scale=0.60]{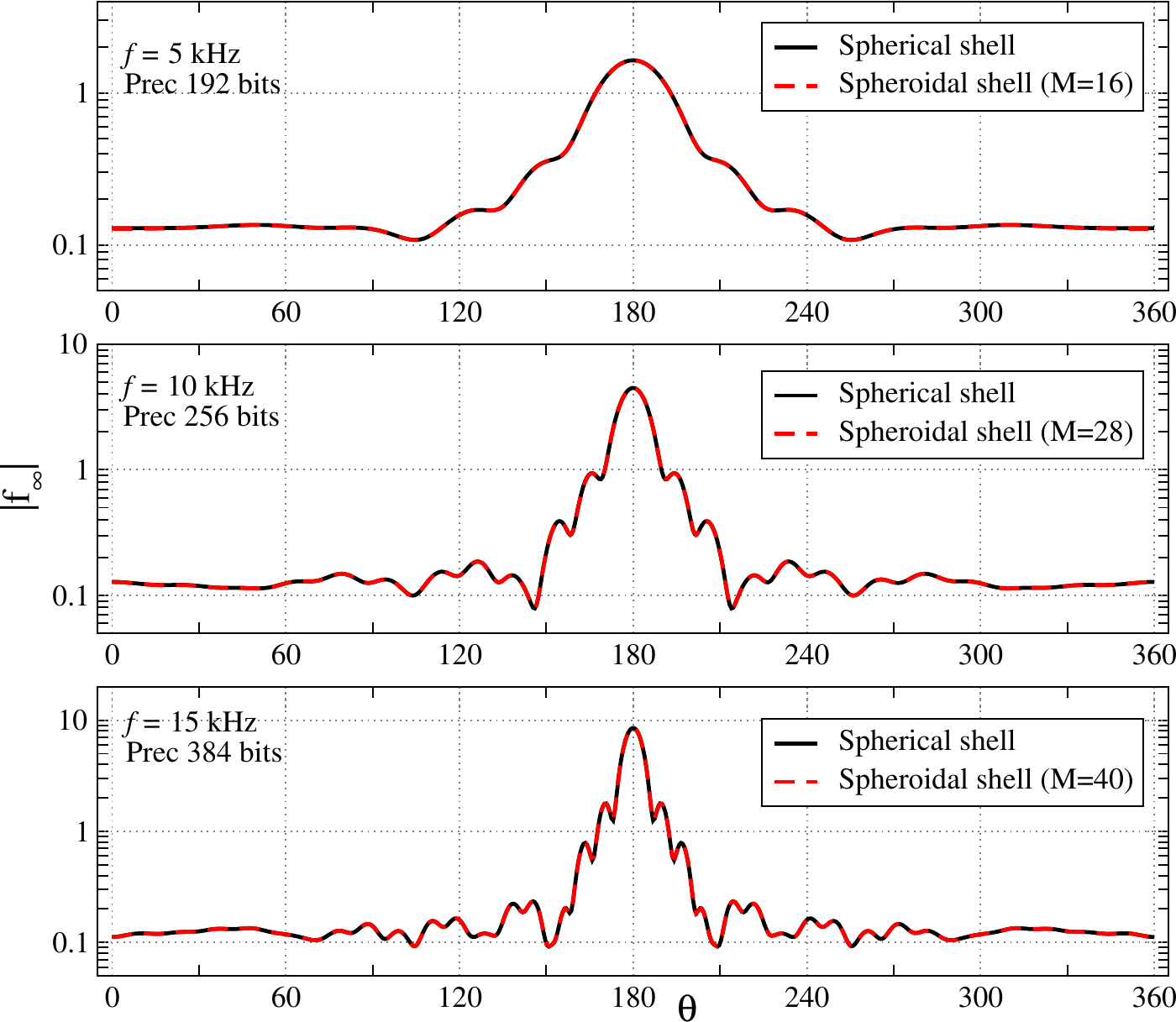}
	\caption{Farfield angular pattern $|f_\infty|$ in terms of the observation angle $\theta$ for a spherical shell (solid lines) and a 
	spheroidal shell (dashed lines) in the geometrical limit when the spheroids tend to spheres. The incidence angle was $\theta_i=180^\circ$.
	\azul{The frequency and precision of the arbitrary precision arithmetic are indicated in the upper left corner of each panel.}}
	\label{fig_verif_geom}
\end{figure}

The results are shown in Figure \ref{fig_verif_geom}, 
\azul{where the frequencies considered are stated in each panel together with the number of bits of precision used by
the arbitrary precision arithmetic ($Bi$ bits of precision imply a floating point number of $\sim \log_{10}(2^{Bi})$ 
significant digits).}
The spherical shell (solid lines) and the spheroidal shell (dashed lines) are in good agreement.
For the spheroidal shell model, the truncation parameter $M$ used in the numerical solution is explicitly indicated
into the graphical window legend. 
Note that a particular $M$-value entails $(M+1)\times (M+1)$ coefficients $A_{mn}$ (cf. Section \ref{num_impl}).

\azul{As can be seen in the successive panels of the figure, a higher frequency requires both a larger truncation
$M$ and more bits of precision to achieve convergence.}
An increase in frequency usually means that more coefficients will be required; so for the $5$ kHz case a
truncation $M = 16$ was sufficient but for $f = 10$ \azul{and $f = 15$} kHz $M = 28$ \azul{and $M= 40$, 
respectively,} were necessary.
Notice that these $M$-values correspond to a converging solution for the external field $p_s$ of Eq. \eqref{pscatt}.
It is not guaranteed that the same external $M$-value will be \azul{enough} for a converging solution \azul{of} 
the internal fields $p_1$ and $p_2$.
In general, the $h \equiv (d/2 )k $ parameter characterize the scattering in a such a way that a higher value of $h$
implies a more oscillatory behavior and consequently more coefficients in the solution are necessary to achieve 
convergence.

\azul{
In this case (spheroidal shell tending to spherical shell) a useful property of the $S_{mn}$ wave functions 
can be used to simplify the error and convergence analysis of the numerical method. As said previously, because 
a spherical shell is isotropic an incidence direction $\eta=-1$ can be used without loss of generality. 
For this particular value all the $S_{mn}$ vanish (a property shared with the opposite angle $\eta=1$) except 
those corresponding to $m=0$.
This property is {\it passed on} to the matricial systems so that the only non-null coefficients $A_{mn}$ are 
the $A_{0n}$ with $n=1,2,...,M$. A fixed truncation $M$ requires then the solution of a single 
$4 (M+1) \times 4(M+1)$ matricial system rather than $M+1$ of them.
For now on, the matrix of the system \eqref{matricial_system} will be indicated as $Q$ for clarity.
}

\begin{figure}[!ht]
	\centering
	\includegraphics[scale=0.70]{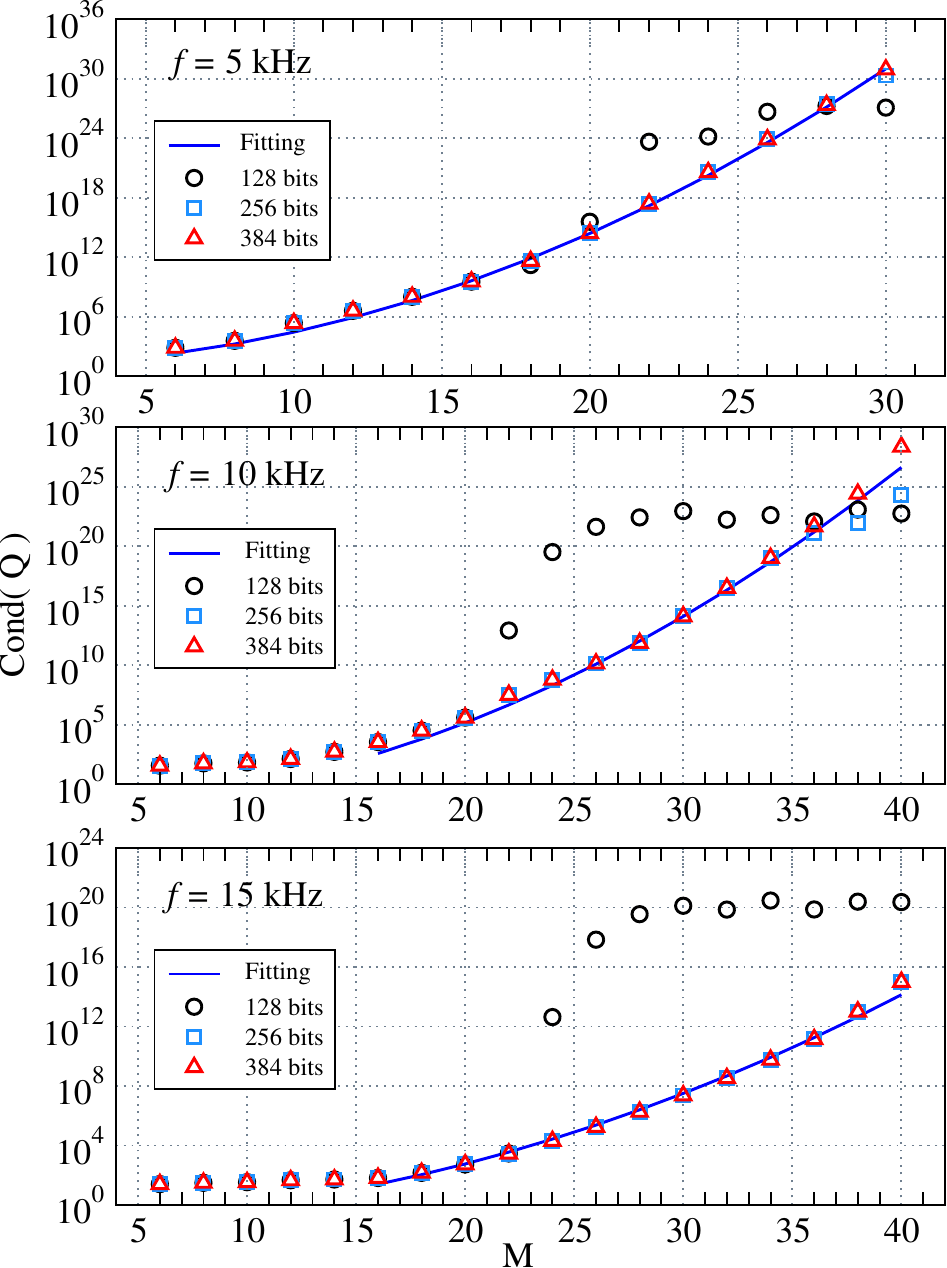}
	\caption{Condition number as a function of truncation $M$ for the matricial system $Q$ corresponding
	to the spheroidal shell tending to a spherical shell.
	For each one of the three frequencies considered (stated in the upper left corner of the panels) 
	precisions of 128, 256 and 384 bits were used. An exponential fitting is shown with solid lines.}

	\label{fig_verif_cond_spherical}
\end{figure}

\azul{
Taking advantage of this feature, the condition number of the (sole) non-null matrix $Q (m=0)$ corresponding to 
a given truncation $M$-value is displayed in Figure \ref{fig_verif_cond_spherical} in log-linear scale.
The top panel shows the case $f = 5$ kHz whereas the middle and bottom ones show $f = 10$ and $f = 15$ kHz. 
Numerical precisions of 128, 256 and 384 bits identified with circles, squares and triangles respectively, 
were considered for each frequency.
}

\azul{
Excluding the (relative) low precision 128 bits curves the behavior in all cases is qualitatively the same; 
an exponential growth of the type $\text{Cond}(Q) \sim e^{M^2}$ as can be checked by a numerical curve fitting 
provided with the ansatz $A e^{B M^2}$, which is shown in solid line in each panel (the values $(A,B)$ resulting
from the fitting were $(13.64, 0.0764), (0.01, 0.0412) $ and $(0.0937, 0.0218)$ for $f=5,10,15$ kHz respectively). 
However, for $f=10$ and $f=15$ kHz the curve fitting was only possible for $M > 15$ so the behavior in the 
entire range of $M$ is maybe more involved.
}

\azul{
Under 128 bit precision and above a certain $M$-value the condition number shows an approximately stalled 
pattern for all frequencies. 
As will be shown below, this is because errors at the spheroidal function evaluation stage. 
Departures between the 256 and 384 bit precision evaluations can also be observed for $f = 10$ kHz but only
for $M \geq 36$.
Remarkably, the condition number decreases as frequency $f$ increases, although this fact is compensated 
because higher frequencies require higher $M$-values to converge (notice that the vertical range in each panel
is different). 
For example, the $f=5$ kHz case converges for $M=18$ where $\text{Cond}(Q) \sim 4\cdot 10^{11}$ whereas the
$f=10$ kHz case reachs convergence for $M=30$ with $\text{Cond}(Q) \sim 1.3\cdot 10^{14}$.
}

\begin{figure}[!ht]
	\centering
	\includegraphics[scale=0.60]{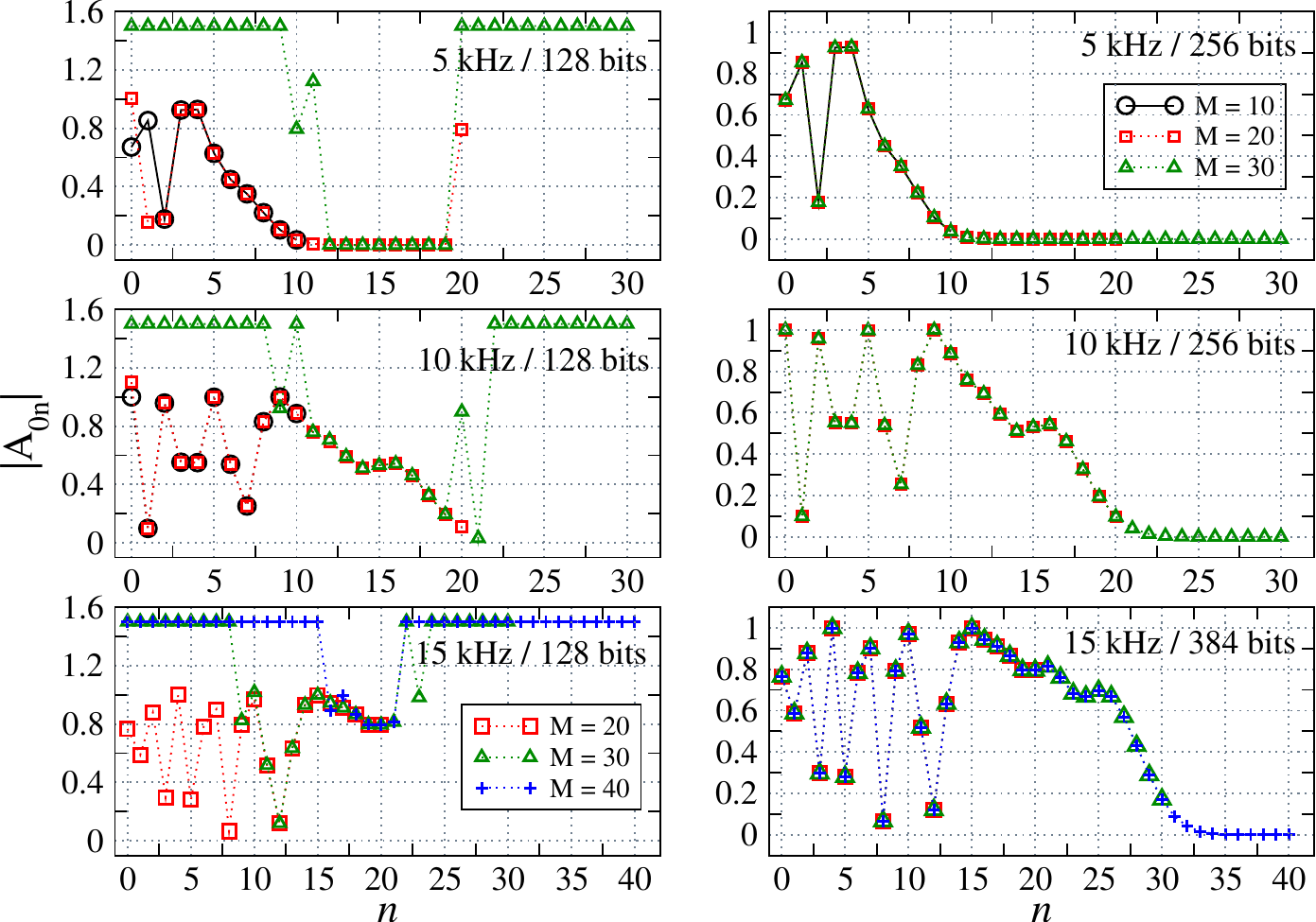}
	\caption{Absolute value of the coefficients $|A_{0n}|$ ($n=0,1,...,M$) as a function of $M$, obtained 
	for the spheroidal tending to spherical shell at the frequencies $f=5,10,15$ kHz and different 
	bit-precisions as listed in the upper right corner of each panel. 
	Low/high precision coefficient computations are showed in the left/right panels.
	Divergent values of $|A_{0n}|$ (occurring in the left panels) are set to a value of 1.5 only for clarity.}
	\label{fig_conv_Amn_spherical}
\end{figure}

\azul{
To throw some light over the convergence of the solution, the absolute value of the $A_{0n}$ coefficients was
evaluated for each frequency considering different sets of $M$-values and numerical precisions.
The six-panel Figure \ref{fig_conv_Amn_spherical} shows the results. 
The different frequencies are considered in ascending order starting from $f=5$ kHz in the top row and the 
left/right columns shows low/high precision evaluations (128 bits/256 or 384 bits, respectively). 
}

\azul{ 
For $f=5$ kHz (top panels) the 256 bits precision evaluation (right) gives a converged solution as can be 
seen since the $M=20$ and the first 21 of the $M=30$ data points display no difference between. 
The left panel shows an interesting effect; up to $M=10$ the eleven coefficients $|A_{0n}|$ have the correct 
values (compare with the first eleven ones in the right side) but suddenly at $M=20$ things goes wrong and 
the $|A_{0 21}|$ takes a incorrect value as well as the first two coefficients but without disturbing the values 
of the intermediate ones $n=2,3,...,20$. 
This situation (correct values for the first coefficients) subsist to $M=17$, so that for this reason a 
converged solution could eventually be calculated with 128 bits of precision if only the first 18 coefficients 
were calculated (the $|A_{0 17}| \sim 10^{-9}$).
In the previous Figure \ref{fig_verif_cond_spherical} (top panel), the condition number of the system at 128
bits was starting to show disturbances just at $M=18$ which can be related to incorrect values in $A_{0n}$ now.
At $M=30$ many of the coefficients take divergent values (up to $10^{27}$) so they are artificially set to a 
value of 1.5 here for the sake of clarity.
}

\azul{ 
For $f=10$ kHz (middle panels) 256 bits of precision still ensures a converged solution but more coefficients 
were necessary (a truncation $M$-value between $M=25$ and $M=30$). 
No converged solution can be obtained with 128 bits because already at $M=20$ errors are noticeable and 
the last coefficients still have non-vanishing values.
The condition number in the middle panel of Figure \ref{fig_verif_cond_spherical} shows departures between 
the solutions with 256 and 384 bits but only after $M=36$ (out of the range shown here).
}

\azul{ 
For $f=15$ kHz (bottom panels) 384 bits were necessary to achieve convergence (obtained for truncation $M$ 
between $M=38$ and $M=40$).
At 128 bits most of the coefficients in the evaluations $M=30$ and $M=40$ acquire wrong values as it is shown 
in the left panel. However, those corresponding to $M=20$ show no appreciable difference with the first 21
correct ones as can be seen from the right panel. Unfortunately, they are not enough for a converged solution.
}

\azul{ 
To quantify errors associated with the solution of the $M+1$ matricial systems \eqref{matricial_system} 
involved in a given scattering problem, a maximum relative error $\varepsilon$ is defined according to
\be
	\varepsilon \equiv \text{Max }_{m=0,1,..,M}
	\left\{ \frac{ | Q^{(m)} x^{(m)} - b^{(m)} | }{ || x || }\right\},
	\label{error_sistema}
\ee
where $Q, b, x$ represent the corresponding matrix, data vector and coefficient vector (solution) 
corresponding to a fixed $m$; the notation $||x||$ stands for the norm of the vector $x$. The absolute
value in the numerator is taken element by element.
}



\begin{figure}[!ht]
	\centering
	\includegraphics[scale=0.40]{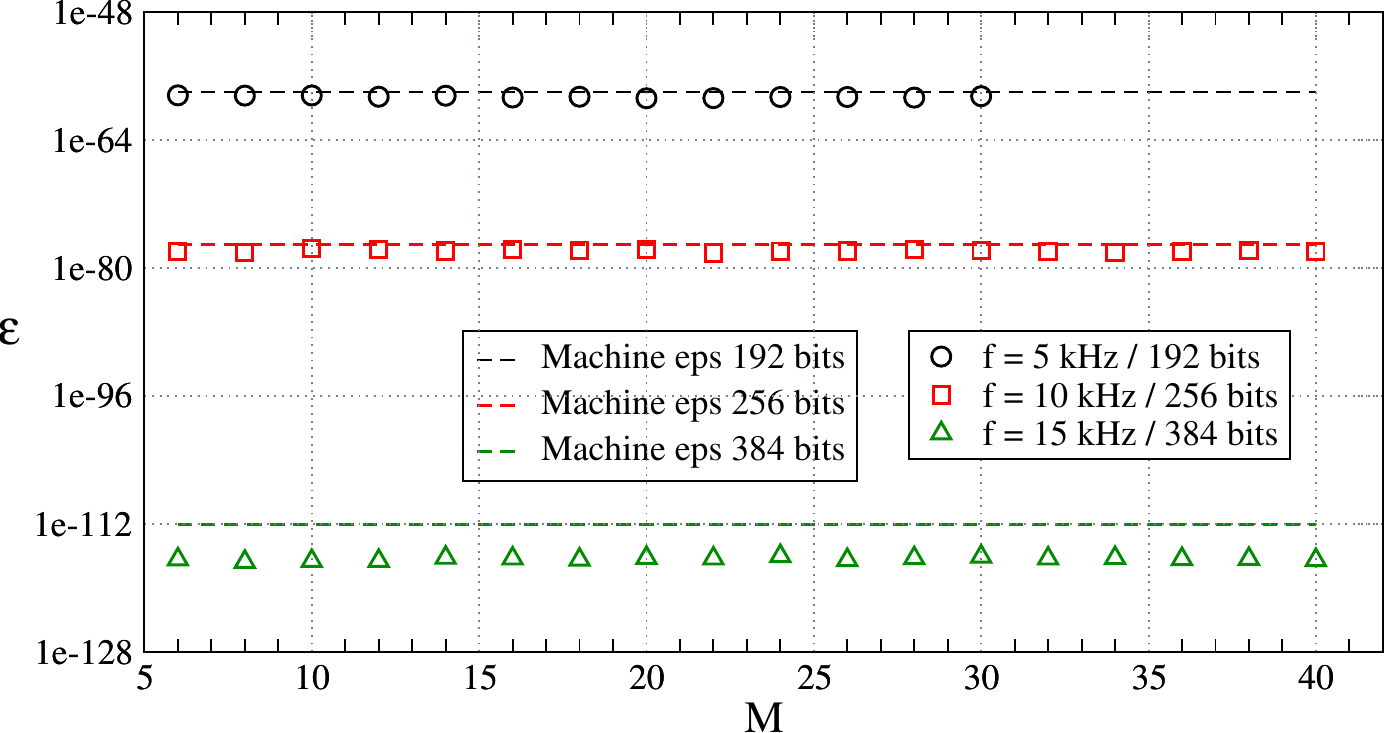}
	\caption{Maximum relative error $ \varepsilon $ as a function of truncation $M$ for the spheroidal
	(tending to sphere) shell for the configurations detailed in the legend plot. Machine epsilons 
	corresponding to the precision used are also plotted with dashed lines.}
	\label{fig_err_spherical}
\end{figure}

\azul{The $\varepsilon$ values obtained from the spheroidal (tending to spherical) shell are displayed
in Figure \ref{fig_err_spherical} in function of the truncation $M$ and according to the configurations
detailed in the plot legend. Additionally, machine epsilon values (unit round-off error) corresponding 
to each bit-precision used were displayed by dashed lines.
Notice that for the previous mentioned reasons only $m=0$ enters in the error evaluation given by
\eqref{error_sistema}.
}

\azul{ 
Even though the condition number is huge (cf. Figure \ref{fig_verif_cond_spherical}) and in addition increases
with the truncation value $M$, this not necessarily imply that the system turns out to be ill-conditioned because
the expected errors in the matricial system assembly are of an order given for the machine epsilon of the actual 
precision of the arbitrary precision arithmetic being used in the calculations.
High precision then makes it possible to tackle matrix systems with a very large condition number.
}

\subsection{Spheroidal geometry by a BEM solution}

In order to verify the behavior of the model in a true spheroidal geometry, a Boundary Element Method (BEM) implementation for the acoustic 
problem of two confocal spheroids was implemented.
The BEM method involves surface integration over the scatterer's boundaries. The usual approach is to consider a discretized version of 
each scattering surface, i.e. a {\it mesh}, composed by simpler elements as triangular o quadrilateral facets, which can be planar or curved.
Then, the boundary integral is converted into a sum of integrations over the mesh elements.

Because the BEM method is an approximation it is expected that its solutions will be only a good approximation to an exact solution.
Nevertheless, if the number of elements in the mesh is greater enough, the method allows for achieving a good agreement with an exact solution.

A usual prescription to ensure the preceding condition is to demand that the length $\ell$ of each segment that constitutes the
mesh verifies a relation
\be
	\ell \leq \lambda / \beta,
	\label{cond_lambda_longitud}
\ee
where $\beta$ is five or six \cite{marburg2002six}. Qualitatively this ensures that the field over the surface is {\it well represented}.

To test the model in a {\it true} confocal spheroidal shell configuration an outer spheroid of $ a_1 = 0.5 $ m and $ b_1 = 0.25 $ m 
and an inner one of $ a_2 = 0.46 $ m and $ b_1 = 0.1552417 $ m were considered. The material properties of the three media involved
are the same ones tabulated in Table \ref{data_table}.

Two meshes representing the external and internal spheroids were built. The external mesh has $ N_E = 21324 $ triangular elements 
whereas the interior one has $N_I = 12380 $.
The maximum segment length in each case was 0.019768 m and 0.0174027 m, respectively, which implies that the maximum wavelengths
$\lambda$ that verify \eqref{cond_lambda_longitud} (in the more strict condition $\beta = 6$) are 0.118608 m and 0.1044162 m.
With these values a maximum frequency $f_\text{max}$ can be calculated for the scattering problem in question, in such a way that 
it is guaranteed that for frequencies less or equal to $f_\text{max}$ all the fields are well represented.

Since there are three sound speeds involved in the problem it follows that the safest situation corresponds to taking the slowest
sound speed and the longest wavelength; that is, to consider the lowest of the maximum frequencies. Taking into account the material properties
from Table \ref{data_table} and the aforementioned meshes, a $f_\text{max} = 2865$ Hz value is obtained.
This does not mean, of course, that scattering evaluation for a frequency greater than the determined $f_\text{max}$ will fail
catastrophically past over that threshold but only that gradual departures are expected as the frequency increases beyond that barrier.

\begin{figure}[!h]
	\centering
	\includegraphics[scale=0.30]{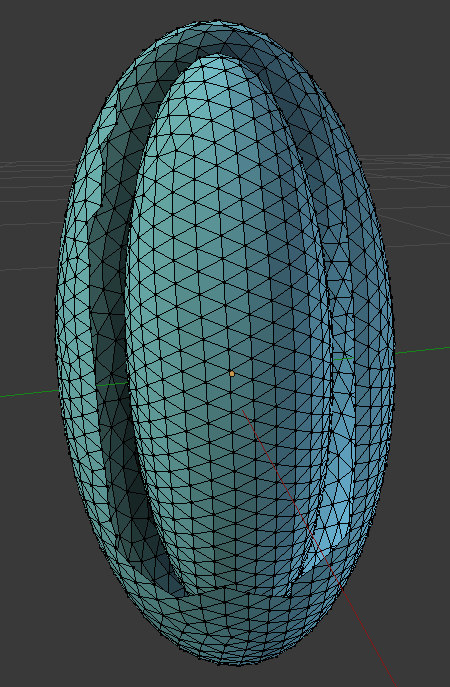}
	\caption{Illustrative mesh for the spheroidal shell. The external and internal spheroid meshes have $N=2248$ and $N=1484$ triangles,
	respectively. The exterior spheroid has some triangles removed so that the presence of the internal one is evident.
	} 
	\label{fig_mesh_spheroids}
\end{figure}

In the Figure \ref{fig_mesh_spheroids} two meshes for the confocal spheroidal setup are displayed. In this case, only for clarity 
purposes, the meshes have a reduced number of elements (2248 and 1484 for the external and internal mesh, respectively) so that individual 
triangles are clearly seen. The external spheroid mesh has also part of its surface removed to allow visualizing the internal one.

\begin{figure}[!htb]
	\centering
	\includegraphics[scale=0.60]{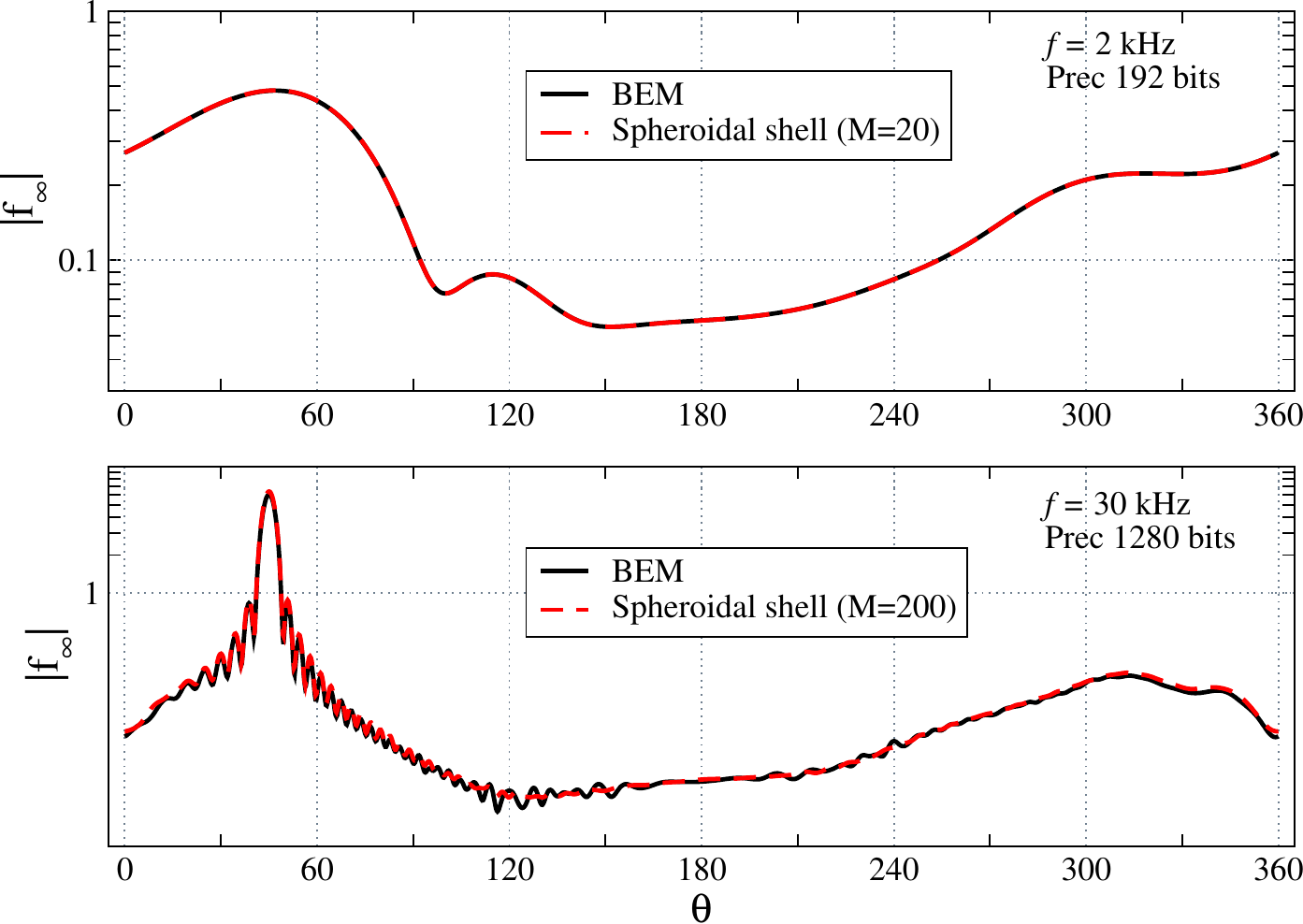}
	\caption{Absolute value of the farfield angular pattern $|f_\infty|$ in terms of the observation angle $\theta$ 
	for the spheroidal shell at frequency $f=2$ kHz (top panel) and $f=30$ kHz (bottom panel). 
	The incidence angle was $\theta_i = \pi / 4 $.
	\azul{The frequency and precision used in the solution are indicated at the right corner of each panel.}
	} 
	\label{fig_sph_bem_2-30}
\end{figure}

\azul{
The top panel of Figure \ref{fig_sph_bem_2-30} shows the resulting angular pattern for the absolute value of the $f_\infty$ 
for $f = 2$ kHz and incidence angle $\theta_i = \pi / 4 $, evaluated with the BEM formulation (solid lines), 
using the meshes $\{ N_E, N_I \}$, and the spheroidal shell model with truncation $ M = 20 $ and precision of
192 bits (dashed lines). Both curves match. 
For the same incidence but $f = 30$ kHz, a frequency value for which the present meshes are clearly insufficient for a 
{\it well represented} scattering, the $|f_\infty|$  is shown in the bottom panel. 
At this frequency, the BEM solution exhibits clear departures from the spheroidal shell solution.
In this case, the spheroidal shell model has required $ M = 200 $ and a precision of 1280 bits to ensure convergence.
The parameters $h$ for this problem are $h_0 = 55.24$, $h_1 = 53.119$ and $h_2 =240.19$, so it is a high frequency
case.
}

\azul{
Given that an incidence angle out of the end-on directions ($\theta=0,\pi$ or equivalently $\eta=1,-1$)
was selected for this numerical verification, all the $M+1$ systems must be taken into account now and
a full matrix of $A_{mn}$ coefficients is implicated.
}

\begin{figure}[!htb]
	\centering
	\includegraphics[scale=0.32]{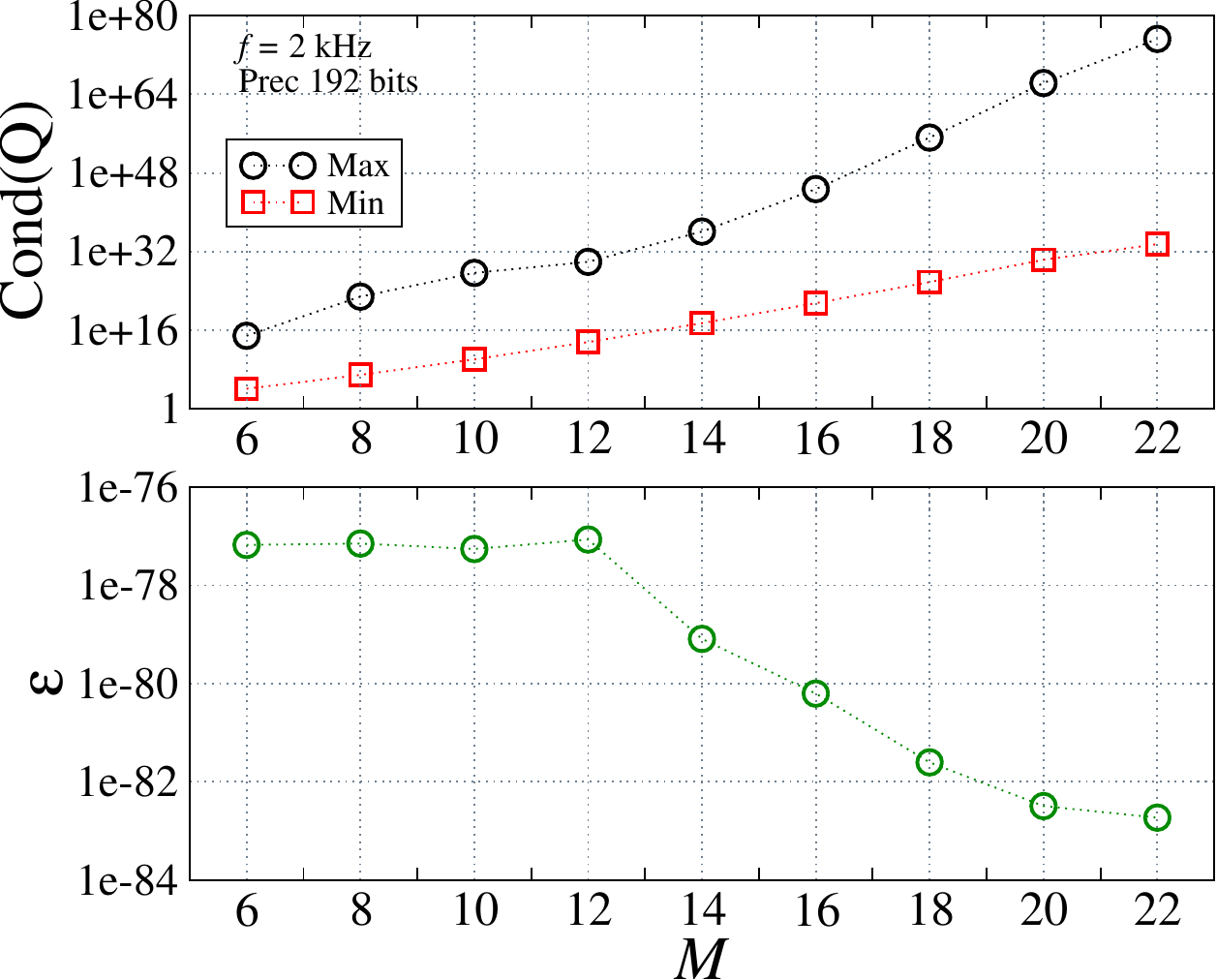} \hspace{1em}
	\includegraphics[scale=0.85]{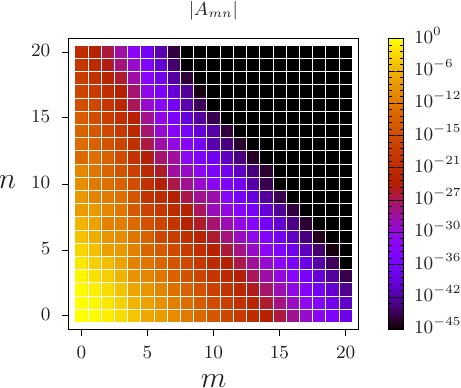}
	\caption{Maximum and minimum condition numbers as a function of $M$ (top panel), maximum relative error
	(bottom) and absolute values $|A_{mn}|$ (truncation $M=20$) for the spheroidal shell system for 
	$f =$ 2 kHz and evaluated at 192 bits of precision.}
	\label{fig_cond_err_bem}
\end{figure}

\azul{
The Figure \ref{fig_cond_err_bem} summarize both the convergence and error analysis for the spheroidal shell
computation in the $f=2$ kHz / 192 bits case. The upper left panel shows the maximum and minimum condition
numbers of the $M+1$ matrices $Q(m)$ ($m = 0, 1,..., M$) involved in the obtained solution at each truncation $M$. 
The bottom left panel displays the maximum relative error in terms of $M$ and the right panel shows the absolute
value $|A_{mn}|$ for every coefficient of the converged solution at $M=20$.
Remarkably, the maximum condition number displays a steeper increase than the minimum (at $M=20$ the difference
between them is of 30 orders of magnitude) despite the fact that the maximum error remains negligible for the 
entire $M$-range as can be verified in the bottom panel.
The $|A_{mn}|$ shows vanishing values towards increasing $m$ and $n$, as would be expected, although the 
vanishing rate seems to be somewhat faster in $m$-direction (fixed $n$).
}

Regarding the computation of the spheroidal shell at $f=30$ kHz ($M=200$ and precision of 1280 bits) showed in the
bottom panel of Figure \ref{fig_sph_bem_2-30}, it should be noted that is a very time consuming problem;
in a 88-core Intel Xeon E5 @2.1 GHz cluster-type environment the computation time was nearly 12 hours.


\subsection{Near field calculation}

Finally, to make use of all the solution coefficients $A, B, C, D$, a nearfield calculation is carried out.
Since this solution will not be compared with a benchmark, material media properties and frequency were selected to produce
an aesthetically more pleasant plot. 
The two confocal spheroids retained the previously used maximum and minimum radius but the frequency was set to
$f = 7.5$ kHz and the material properties were according to Table \ref{data_table_nearfield}.
The amplitude and incidence angle of the incident wave were $ p_0 = 1 $ and $ \theta_i = \pi / 4 $, respectively.

\begin{table}[h]
	\begin{center}
	\begin{tabular}{ccc}
	\hline
	\hline 
	\vspace{-0.8em}\\
	Medium & $c$ (m s$^{-1}$) & $\rho$ (kg m$^{-3}$) \\
	\hline
	\vspace{-1em}\\
	0 (water) & 1477.4 & 1026.8 \\
	1 & 3 $c_0$ & 1.25 $\rho_0$ \\
	2 & 1.5 $c_0$ & 2.25 $\rho_0$ \\
	\vspace{-1em}\\
	\hline
	\hline
	\end{tabular}
	\end{center} 
	\caption{Material properties (sound speed $c$ and density $\rho$) for nearfield evaluation.}
	\label{data_table_nearfield}
\end{table}

\begin{figure}[!htb]
	\centering
	\includegraphics[scale=0.4]{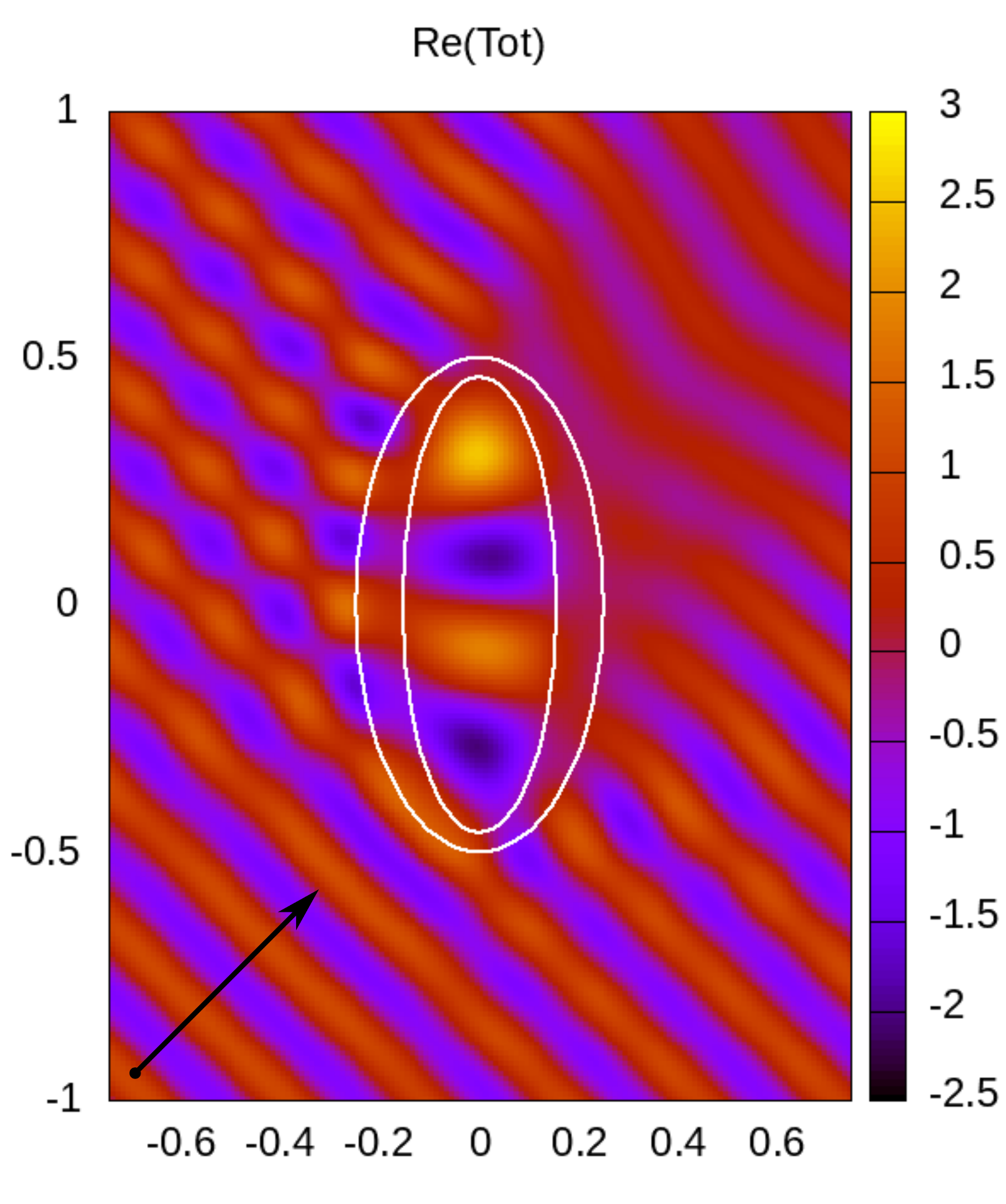}
	\hspace{1em}
	\includegraphics[scale=0.4]{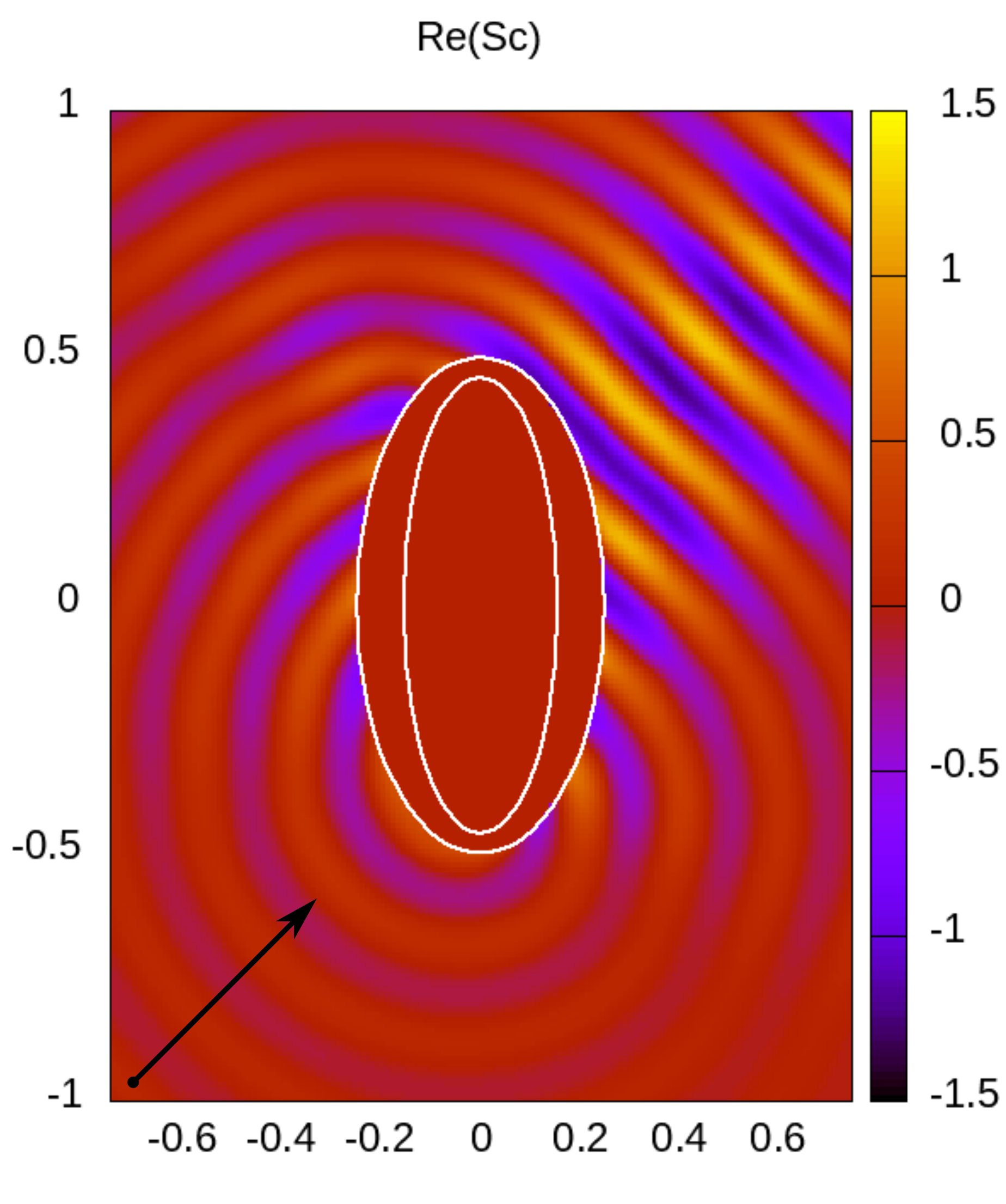}
	\caption{Real parts of the total (left) and the scattered (right) pressure field evaluated over the plane $ y = 0 $ in the
	nearfield region of the spheroidal shell. The incident field considered has amplitude $p_0 = 1$ and incidence angle 
	$\theta_i = \pi /4$ (the incidence direction is indicated by an arrow).
	} 
	\label{fig_nearfield}
\end{figure}

In Figure \ref{fig_nearfield} (left panel) the real part of the total field is shown in the interior of each spheroid and also
in its surroundings, evaluated over the plane $y = 0$. The right panel of the figure exhibits the real part of the scattered field 
which exists only in the exterior to the external spheroid. The incidence direction is indicated by an arrow in both panels.
The solution of the field in all regions required $ M = 40 $ and the parameters $h$ were $h_0 = 5.98$, $h_1 = 1.99$  and $ h_2 = 3.98$
so it is, indeed, an intermediate frequency scattering problem.

The total field displays no continuity problems or artifacts when crossing each one of the spheroid's boundaries, indicated
on the figure by ellipses. This constitutes and indirect verification of the solution since the field in points located near 
each boundary but at opposite sides have been calculated through a different set of coefficients but they show, 
however, due to continuity, no abrupt changes in the field values.
A shadow zone in the opposite side of the incidence is noticeable as well as an intense field value zone in the innermost spheroid.

The scattered field displayed in the right panel seems to correspond to a spherical source located in some point at the bottom of the 
shell, modified by the presence of the incident field in the previously mentioned shadow zone. 
Of course, this is expected because a total field near zero corresponds to $ p_s \sim -p_i $, according to Eq. \eqref{pressure_out}.

\section{Conclusions}
\label{conclusions}

A model to calculate the external and internal fields in the case of two confocal spheroids (an spheroidal shell)
was presented. The required spheroidal wave function evaluation is carried out by using a previously published code, modified
and optimized by the author to take advantage of parallel execution and also to strengthen high frequency calculations.

Numerical verifications against spherical shell and BEM solutions under certain circumstances allow to infer that the model
adequately solves the scattering problem in a wide frequency interval.
It must be noted that for a very high frequency regime the spheroidal wave function evaluations are computationally expensive 
but in this situation it is very likely that their asymptotic expressions can be used to alleviate that burden.
However, it should be ensured that differences with exact evaluations under this regime would be negligible.

The procedure used for the numerical calculation of the coefficients follows closely the classical one for spherical coordinates
but instead must be solved by truncation. 
The scattering problem for a multilayered spheroidal shell with three or more surfaces can be worked out following the same lines.
\azul{Analysis performed on the condition number of the resulting matricial system and in the convergence of the solution
showed that the precision used by the arbitrary precision arithmetic system is a key element in both evaluation of the
spheroidal functions and coefficient's calculation.}

If care is taken into account for determining the convergence conditions while ensuring the correct evaluation of
the spherical wave functions, the presented model for scattering from confocal prolate spheroids can be added
to the toolkit of exact solutions of the computational physicist devoted to acoustics.

\section*{Acknowledgments}
\label{acknow}
The author wish to thank to Dr. Juan D. Gonzalez and M. Sc. Silvia Blanc for their valuable comments about mathematical 
and physical aspects of the problem as well as his scrupulous reading of the manuscript.

\clearpage

\end{document}